\documentclass[pra,twocolumn, floatfix, footinbib,superscriptaddress, longbibliography]{revtex4-2}
\usepackage[utf8x]{inputenc}
\usepackage[english]{babel} 
\usepackage[english]{babel} 
\usepackage[colorlinks=true,bookmarks=false,citecolor=blue,urlcolor=blue]{hyperref}

\usepackage{amsmath,amssymb,esint}
\usepackage{graphicx}
\usepackage{graphics}
\usepackage{verbatim}
\usepackage{physics}
\usepackage{bbold}
\usepackage{color}
\usepackage[english]{babel}
\usepackage{bm}
\usepackage{natbib}
\usepackage[normalem]{ulem}
\usepackage{epstopdf}
\usepackage{multirow}
\usepackage{ dsfont }
\usepackage{tikz}
\DeclareMathAlphabet\mathbfcal{OMS}{cmsy}{b}{n}

 \usepackage{booktabs}
 \usepackage{multirow}
\usepackage{ragged2e}

\usepackage{amsmath}
\usepackage[colorinlistoftodos]{todonotes}

\newcommand{\bee}{\begin{equation}}
\newcommand{\ene}{\end{equation}}

\newcommand{\bea}{\begin{eqnarray}}
\newcommand{\ena}{\end{eqnarray}}

\newcommand{\red}[1]{{\color{red}#1}}

\graphicspath{{pics/}}

\let\vec=\mathbf

\renewcommand {\phi}{\varphi}

\begin{document}
\title{Symmetry analysis and multipole classification of eigenmodes in electromagnetic resonators for engineering their optical properties }
\author{Sergey Gladyshev}
\affiliation{ITMO University, St. Petersburg 197101, Russia}
\author{Kristina Frizyuk}
\affiliation{ITMO University, St. Petersburg 197101, Russia}

\author{Andrey Bogdanov}
\affiliation{ITMO University, St. Petersburg 197101, Russia}

\date{January 2019}


\begin{abstract}
The resonator is one of the main building blocks of a plethora of photonic and microwave devices from nanolasers to compact biosensors and magnetic resonance scanners. The symmetry of the resonators is tightly related to their mode structure and multipole content which determines the linear and non-linear response of the resonator. Here, we develop the algorithm for the classification of eigenmodes in resonators of the simplest shapes depending on their symmetry group. For each type of mode, we find its multipole content. As an illustrative example, we apply the developed formalism to the analysis of dielectric triangular prism and demonstrate the formation of high-Q resonances originated due to suppression of the scattering through the main multipole channel. The developed approach one to engineer, predict, and explain scattering phenomena and optical properties of resonators and meta-atmos basing only on their symmetry without the need for numerical simulations, and it can be used for the design of new photonic and microwave devices.            
      
%Using the term symmetric resonator analysis we mean the problem of classification and determine the multipole composition of the resonator eigenmodes depending on its symmetry group. The solution to this problem allows a deeper study of the nature of many optical effects.  
%We classify the eigenmodes of the simplest forms of resonators according to their irreducible representations. In particular ..... %cone, % $D_{\infty v}$cylinder, % $D_{\infty h}$	and triangular %prism  %$D_{3h}$ are considered. 
%Multipolar moments of symmetry of the analysis are determined for each type of mod. This allows to identify the radiation patterns and the distribution of the near field. 
%%A symmetry analysis is performed for multipoles \red{until octupoles.}%of order $n<3$. 
\end{abstract}

\maketitle

\section{Introduction}

Dielectric particles with high refractive index recently recommended themselves as a very prospective tool for light manipulation at the nanoscale. Their unique optical properties appear owing to Mie resonances, which can be excited even in the visible or near-infrared ranges for nanoscale particles~\cite{bohren2008absorption,kuznetsov2016optically, arbabi2015dielectric}. Today dielectric particles with high refractive index have shown their for second~\cite{smirnova2018multipolar, kruk2017nonlinear, gili2016monolithic,makarov2017efficient} and third ~\cite{smirnova2016multipolar, shcherbakov2015nonlinear, grinblat2016enhanced} harmonic generation, sensing~\cite{bontempi2017highly}, efficiency light localization ~\cite{kivshar2017meta}, enhancement of the outcoupling radiation~\cite{Kryzhanovskaya2018,Polubavkina2016}, excitation of guided modes~\cite{Sinev2017,krasnok2018all,li2015all,picardi2019experimental}, heating~\cite{zograf2017resonant}, enhancement of Raman scattering~\cite{frizyuk2018enhancement} etc. Along with low absorption in the visible and infrared ranges, an essential advantage of dielectric particles on their plasmonic counterpart is a pronounced magnetic response.  Interplay between Mie  resonances results in many beautiful phenomena like directional light scattering ~\cite{geffrin2012magnetic, shamkhi2019transverse, jin2010metamaterial, liu2018generalized, person2013demonstration, hancu2013multipolar}, anapole and invisibility~\cite{rybin2015switching, miroshnichenko2015nonradiating, luk2017suppression, alu2008multifrequency, alu2007cloaking}, supercavity mode ~\cite{rybin2017high, koshelev2019nonrad}, electromagnetically-induced-transparency~\cite{chiam2009analogue} etc. All these effects can be easily explained in terms of multipole expansion formalism~\cite{alaee2018electromagnetic, grahn2012electromagnetic, Evlyukhin2016}. The multipole formalism is very natural and convenient for small particles, when only several first resonances are essential. Thus, the Kerker effect is explained by constructive interference of electric and magnetic dipole resonances~\cite{kerker1983electromagnetic}, supercavity mode can be explained as cancelation of the dominant multipole moment of the mode. The selection rules for nonlinear harmonic generation in nanoantennas can be easily formulated in terms of multipole moments~\cite{frizyuk2019second, frizyuk2019second1}.  Multipole formalism allows us to better study the properties of the eigenmodes of plasmon nanoparticles \cite{PhysRevB.89.165429,PhysRevB.96.045406}.

%The eigenmode spectrum of a resonator is the most vital characteristic determining optical properties. To describe the properties of small optical resonators - when only several first resonances dominate, it is convenient to use the method of the multipole decomposition. The use of such formalism simplifies the analysis of the scattering features and nonlinear processes. %Using this formalism, the radiation pattern, selection rules for the generation of harmonics, and scattering features are described quite simply. 

However, a one-to-one correspondence between the modes and multipoles can be set only for spherical resonators (spheres, core-shell particles, voids, etc). It occurs because of the angular part of the modes can be separated from the radial part of the wavefunction only for objects with spherical symmetry. Pure spherical resonators in a homogeneous environment are found in colloidal solutions or atmospheric physics but for integrated photonics, this is a rather unique case due to the inability to fabricate a spherical resonator by lithographic methods. For non-spherical resonators, even for a sphere on a substrate, the eigenmodes are contributed by an infinite series of the multipole moments, which depends on the symmetry of the resonator.  The knowledge of the specific multipole composition of the eigenmodes depending on the resonator's symmetry is very important for engineering photonics structures with on-demand optical properties.

% The symmetry analysis allows to classify the eigenmodes depending in the symmetry if the resonator and predict the specific multipole series for each type of the mode.      

%using the symmetry analysis allows to predict the  the symmetry of the resonator and its modes 
%
%it is possible to predict all allowed multipole   
%
%For an arbitrary shape of the resonator, all the coefficients of the multipole series are not zero in the general case but in virtue of the resonator's symmetry some multipoles can be forbidden.  if the resonator has symmetries  
%
% Therefore, multipole analysis of non-spherical resonators is very important problem of modern nanophotonics.

In this work, we provide a comprehensive multipole analysis for non-spherical resonators. Using the group-theoretical approach, we classify the mode types in resonators of different shapes and find the specific multipole series inherit to each type of the modes.  In particular, we consider the resonators of the symmetry group of the cylinder, cone, cube, triangular and quadratic prisms, and chiral resonator. As an illustrative example, we apply the developed formalism to the dielectric triangular prism and analyze the evolution of the multipole content of the modes with the geometrical aspect ratio of the prism. We found a supercavity mode for the specific aspect ratio, when the radiative losses through the dominant multipole channel disappear and the quality factor demonstrates giant growth.

 \begin{figure}[t!]
\includegraphics[width=0.9\linewidth]{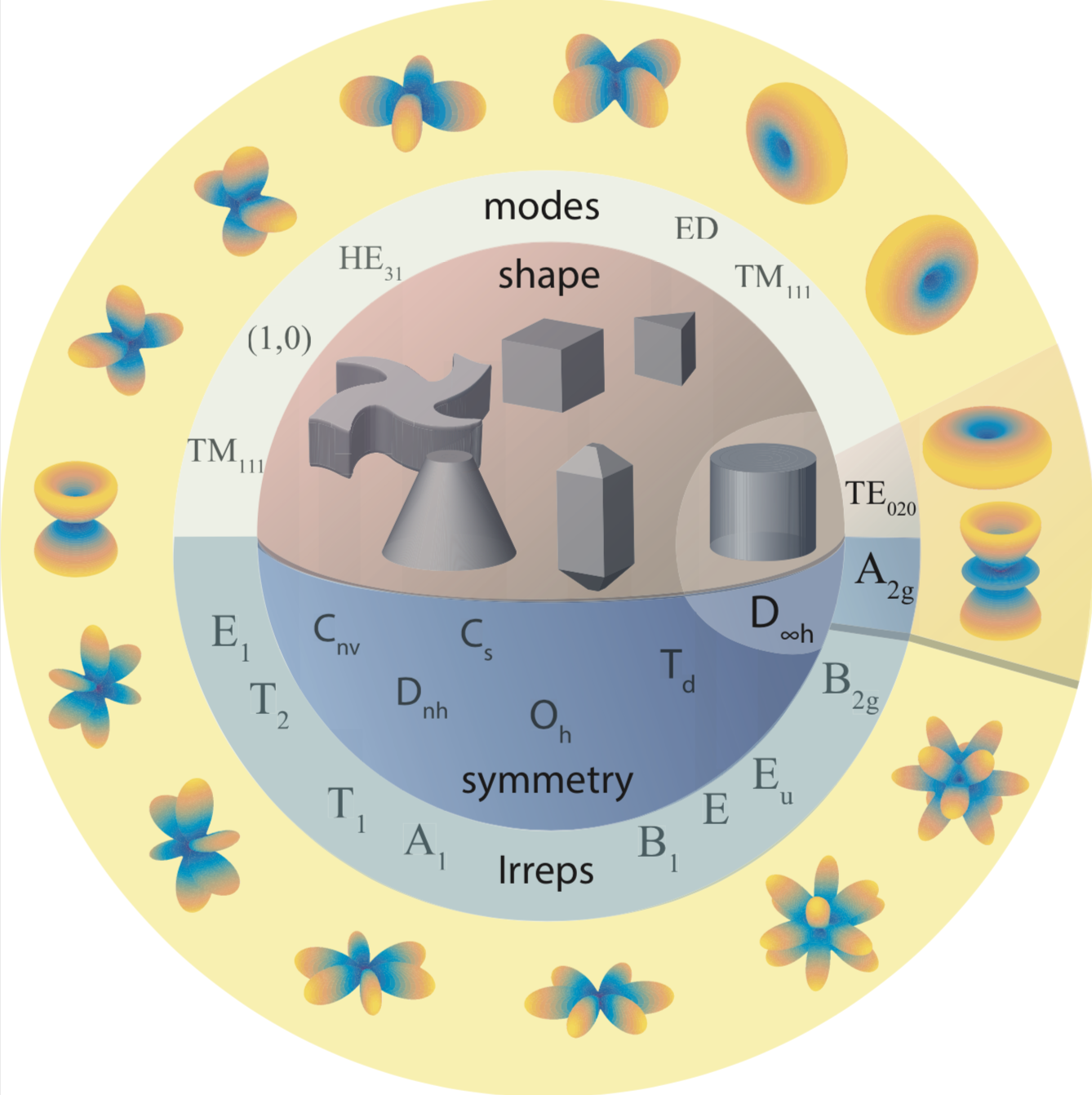}%
\caption{Sketch showing the main idea of the paper.  The resonators of different shapes can be classified according to their group symmetry and their eigenmodes are classified according to the irreducible representations of the resonator's symmetry group. Each irreducible representation can be characterized by a set of vector spherical harmonics defining the multipole content of each mode.}
\end{figure}

\section{ Analisys of multipole content}
  
%Eigenmodes of a homogeneous dielectric sphere known as quasi-normal modes~\cite{1} or resonant states~\cite{1} are satisfy to the Helmoltz equation with open boundary condition (Sommerfeld boundary condition):
%\begin{eqnarray}
%\Delta \vec{E} + \varepsilon(\mathbf{r})\left(\frac{\omega}{c}\right)^2 \vec{E}=0, \\
%\varepsilon(\mathbf{r})=\left\{
%   \begin{array}{cc} % brackets may be (...), [...], \{...\}, or left out
%      \varepsilon & \text{when} \ |\mathbf{r}|\leq R  \\
%      1 &  \text{when} \ |\mathbf{r}|> R  
%   \end{array}
%\right. .
%\end{eqnarray}  
The eigenmodes of a homogeneous dielectric sphere $\mathbfcal{E}_s(\vec{r})$ known as quasi-normal modes~\cite{lalanne2018light} or resonant states~\cite{doost2014resonant} are contributed by the single multipole moments. In other terms, each resonant state contains only one vector spherical harmonic (VSH) and $\mathbfcal{E}_s(\vec{r})\sim\mathbf{W}_s(\mathbf{r})$. Here, index $s$ is a set of the indices $\{t, p, m, l\}$ encoding the VSH $\mathbf{W}_{t pml}$. Index $l=0,1,2, ...$ is the total angular momentum quantum number, and $m=0,1,...,l$ is the absolute value of the projection of the angular momentum (magnetic quantum number).
 Index $p=\pm1$ defines the parity of $\mathbf{W}_s$ with respect to reflection from the $xz$-plane ($\varphi\rightarrow-\varphi$). If $p=\pm1$ then $\mathbf{W}_s\rightarrow\pm\mathbf{W}_s$ under reflection form the $xz$-plane. Let us note that this definition of parity differs from that given in Ref.~[\onlinecite{bohren2008absorption}].
 Index $t$ denotes the polarization of the VSH and parity under inversion transformation. We put $t=(-1)^{l+1}$ for the magnetic VSHs (usually denoted as $\mathbf{M}$) and we put $t=(-1)^{l}$ for the electric VSHs (usually denoted as $\mathbf{N}$) (see, e.g., Ref.~[\onlinecite{bohren2008absorption}]). More detailed information about VSHs is provided in Appendix \ref{dvsh}.   All these states form a complete set and, therefore, any eigenstate of a non-spherical resonator $\mathbf{E}_q$ can be expanded into the series of $\mathbfcal{E}_s(\vec{r})$~\cite{muljarov2011brillouin}:         
\begin{equation}
\label{eq:expansion}
\mathbf{E}_q(\mathbf{r})=\sum_s C_{s}^q\mathbfcal{E}_s(\mathbf{r}).
\end{equation}
Therefore, the multipole content of $\mathbf{E}_q(\vec{r})$ is completely determined by the set of resonant states $\mathbfcal{E}_s(\vec{r})$ contributing to this mode. Strictly speaking expansion \eqref{eq:expansion} also contains longitudinal (non-solenoidal) harmonics $\mathbf{L}$~\cite{lobanov2019resonantstate}. This fact does not affect the further symmetry analysis because the harmonics $\vec N$ and $\vec L$ with the same indices have the same symmetry with the respect to the transformations of O(3) group ~\cite{lobanov2019resonantstate}. Some coefficients in Eq.~\eqref{eq:expansion} can vanish in virtue of the symmetry of the mode  $\mathbf{E}_q(\vec{r})$, which in turn is determined by the symmetry of the resonator. There are two ways to find non-vanishing coefficients. 

 The first way is to use the fact that eigenmodes are transformed according to irreducible representations of the symmetry group of the resonator~\cite{ivchenko2012superlattices}. Thus, we have the one-to-one correspondence between the irreducible representations and the types of the eigenmodes of the resonator, and each eigenmode corresponding to a certain irreducible representation consists of only those VSHs, which are transformed according to the same irreducible representation. The irreducible representations are well known for all point symmetry groups in three-dimensional space~\cite{irrep}. To find the VSHs composing the basis of a certain irreducible representation it is possible to use the projection operator \cite{Tinkham, Xiong:20} which helps to bring the transformation matrices to a block diagonal form. The transformation matrices are obtained by substituting the particular angles into D-Wigner matrices (Appendix \ref{trans}). To avoid dealing with D-Wigner matrices we can use the fact that electric VSHs are transformed as scalars harmonics under rotations and inversion in O(3) and magnetic VSHs are transformed in an opposite way under inversion, i.e. as pseudo-tensors. For some particular groups, the symmetry behavior of VSHs is analyzed in literature~(see, e.g., \cite{frizyuk2019second1,Xiong:20}).

%   ~\cite{spice, PhysRevB.98.165110, doi:10.1002/9783527695799.ch12} 
% 
%Another way to find the basis of a certain irreducible representation in terms of VSHs is to use the fact   
%
%
%
%However, the simple way is to compare the symmetry behavior of the vector and scalar spherical harmonics.  
%  we can note that every point group is a subgroup of the O(3). This means that the harmonics with the same $l$ are transformed through each other. One of the ways 

 The second way to find non-vanishing coefficients in Eq.~\eqref{eq:expansion} is based on the resonant state expansion scheme~\cite{doost2014resonant}. {This way doesn't require the irreducible representation formalism to find the multipolar content of each mode. %Moreover, it allows us to consider some exceptional cases more carefully, even then the previous method fails.} %cylinder m=0
 Let us focus on this way more detailed.

%\red{In the following considerations we exploit }the fact that the eigenmodes are transformed according to irreducible representations of the symmetry group of the structure ~\cite{ivchenko2012superlattices}, \red{so we obtain a connection between the symmetry of the structure and the symmetry of the eigenmode. Further, the nature of this connection will be explained in detail.}

Any finite dielectric resonator can be represented as a perturbation of the circumscribed sphere (see Fig.~\ref{perp}). The resonant state of a non-spherical resonator $\mathbf{E}_q(\vec{r})$ satisfy the Helmholtz equation with open boundary condition (Sommerfeld boundary condition):   
\begin{equation}
\nabla \times \nabla \times \mathbf{E}_{q}=[\varepsilon(\mathbf{r})-\Delta \varepsilon(\mathbf{r})] \frac{\Omega_{q}^{2}}{c^{2}} \mathbf{E}_{q}.
\end{equation}
It follows from the generalized Brillouin-Wigner perturbation theory~\cite{brillouin1932problemes} that  coefficients $C_{s}^q$ satisfy the following linear system:
\begin{equation}
\frac{1}{\omega_{s}} \sum_{s'}\left[\delta_{s s'}+V_{s s'}\right] C_{s'}^{q}=\frac{1}{\Omega_{q}} C_{s}^{q}.
\end{equation}          
Here $\omega_s$ and $\Omega_q$ are the complex eigenfrequency corresponding to  $\mathbfcal{E}_s(\vec{r})$ and $\vec{E}_q(\vec{r})$, respectively. The matrix elements  $V_{s s'}$ defined as   
\begin{align}
&V_{s s' }=-\frac{1}{2}{\int{ \Delta\varepsilon({\bf r}) \mathbfcal{E}_s(\mathbf{r}) \cdot \mathbfcal{E}_{s'}(\mathbf{r})}}\dd {\bf r} \\
\label{eq:matrix_element}
&V_{s s' }\sim \int{ \Delta\varepsilon({\bf r}) \mathbf{W}_s(\mathbf{r}) \cdot \mathbf{W}_{s'}(\mathbf{r})}\dd {\bf r}
\end{align}
are responsible for the coupling between different resonant states with indices $s$ and $s'$ in the perturbed resonator.

  According to the selection rule theorem for matrix elements~\cite{mechanics1977landau}, such an integral over the  nanoparticle's volume is nonzero only if the integrand is invariant with respect to all transformations of the particle's symmetry group. That is why we are interested only in symmetry behavior of the vector spherical harmonics. Harmonics $\vec{W}_{t pml}$ and $\vec{W}_{t' p' m' l'} $ belong to the similar mode,  if the integral in Eq.~\eqref{eq:matrix_element} is non-zero. {The product of VSHs can be expanded into a series of scalar tesseral spherical harmonics $\psi_{p m l}$ %(see Appendix \ref{trans})
   as follows~\cite{frizyuk2019second}:  }
  \begin{equation}\label{comp} 
 \vec{W}_{t p m l }\cdot \vec{W}_{t' p' m' l'}  =\!\!\!\!\!\!\!\!  \sum_{
    \begin{matrix} % or pmatrix or bmatrix or Bmatrix or ...
       {\scriptstyle m''=|m\pm m'|} \\
       {\scriptstyle |l-l'|\leqslant l'' \leqslant l+l'}\\
    \end{matrix}
 }\!\!\!\!\!\!\!\! \mathbb{1}_{tt'}^{l''}  a_{p''m''l''}(r) \psi _{p''m''l''}.
    \end{equation}
Here $\psi_{pml} $ are the functions invariant with respect to symmetry transformations of the resonator, 
index $p''=pp'$, and $a_{p''m''l''}(r)$ is the function of radius-vector $r$. Indicator $\mathbb{1}_{tt'}^{l''}$ shows that the sum is taken over even $l''$ if $tt'=1$ and the sum is taken over odd $l''$ if $tt'=-1$. %This expansion is based on symmetry behavior of vector spherical functions.
      
    %  \blue{It is not clear, why do we need the next paragraph here.}
      
    %   \red{For every scalar spherical function the transformation law is known, i.e. they are transformed by the Wigner D-Matrixes \cite{dwigner}, however, we can often reveal that some of them are invariant under certain symmetry group transformations by simply looking at its image (see Fig.~\ref{fig:harmonics})}. The transformation law of the scalar and  vector  spherical functions is provided in Appendix \ref{trans}. 

In order to find all $\vec{W}_{tpml}$ contributing to a certain mode one can follow the following algorithm:
\begin{enumerate}
\item  Find the functions $\psi_{p''m''l''}$, which are invariant with respect to symmetry transformations of the resonator. The simplest way to this is just have a look at images of the scalar spherical harmonics (see Fig.~\ref{fig:harmonics}) and an image of $\Delta\varepsilon$. 
\item Take an arbitrary function  $\vec{W}_{tpml}$. It is convenient to consider the function with the lowest indices. For instance, a function corresponding to magnetic or electric dipole.
\item Find harmonics $\vec{W}_{t'p'm'l'}$ coupled to $\vec{W}_{tpml}$ using the following relations:
	\begin{enumerate}
		\item $t' = (-1)^{l''}t $;
		\item $p'= p''/ p$;
		\item  search all $l'$ such as $|l-l'|\leq l'' \leq l+l'$;  
		\item  search all $m'$ such as $m'\leq l'$ and $m''=|m \pm m'|$;  
	\end{enumerate}
\end{enumerate}
Repeating this procedure for different initial functions $\vec{W}_{tpml}$ and $\psi_{p''m''l''}$ we will find that all VSHs are divided into groups. Each of these groups corresponds to some irreducible representation. Therefore, the proposed algorithm gives a way to classify the modes of resonators and find the multipole series contributing to each mode type.    

\begin{figure}
     	\includegraphics[width=0.95\linewidth]{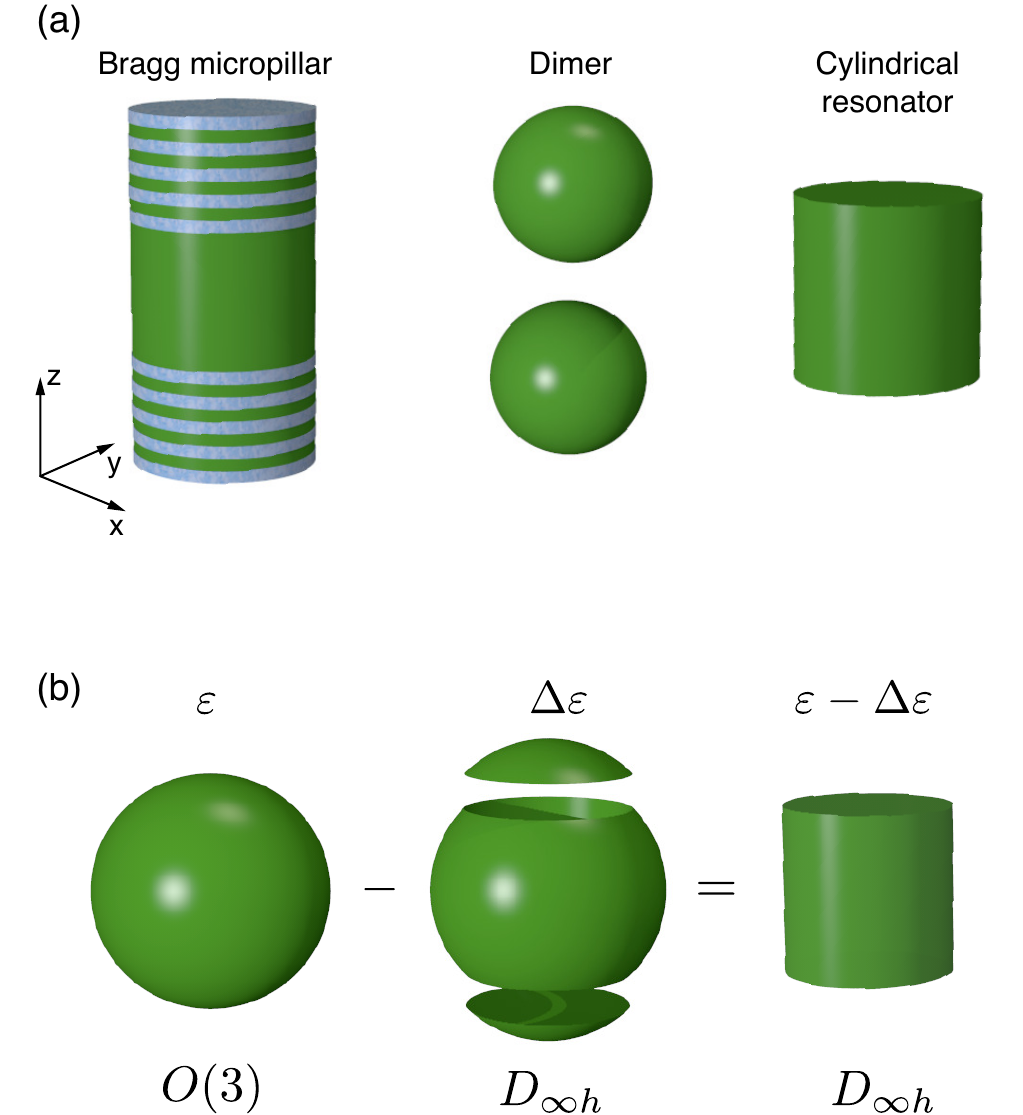} 
     	\caption{(a) Examples of different resonators of the same symmetry group $D_{\infty h}$. (b) Transformation the sphere [$O(3)$ group] with dielectric susceptibility  $\varepsilon$ to the resonator ($D_{\infty h}$ group) with  dielectric susceptibility $\varepsilon + \Delta \varepsilon$. }
     	\label{perp}
\end{figure}

 \section{Results of multipole analysis}
 \subsection{Cylindrical resonators}
 
 \begin{figure*}[t!]
	\includegraphics[width=1.\linewidth]{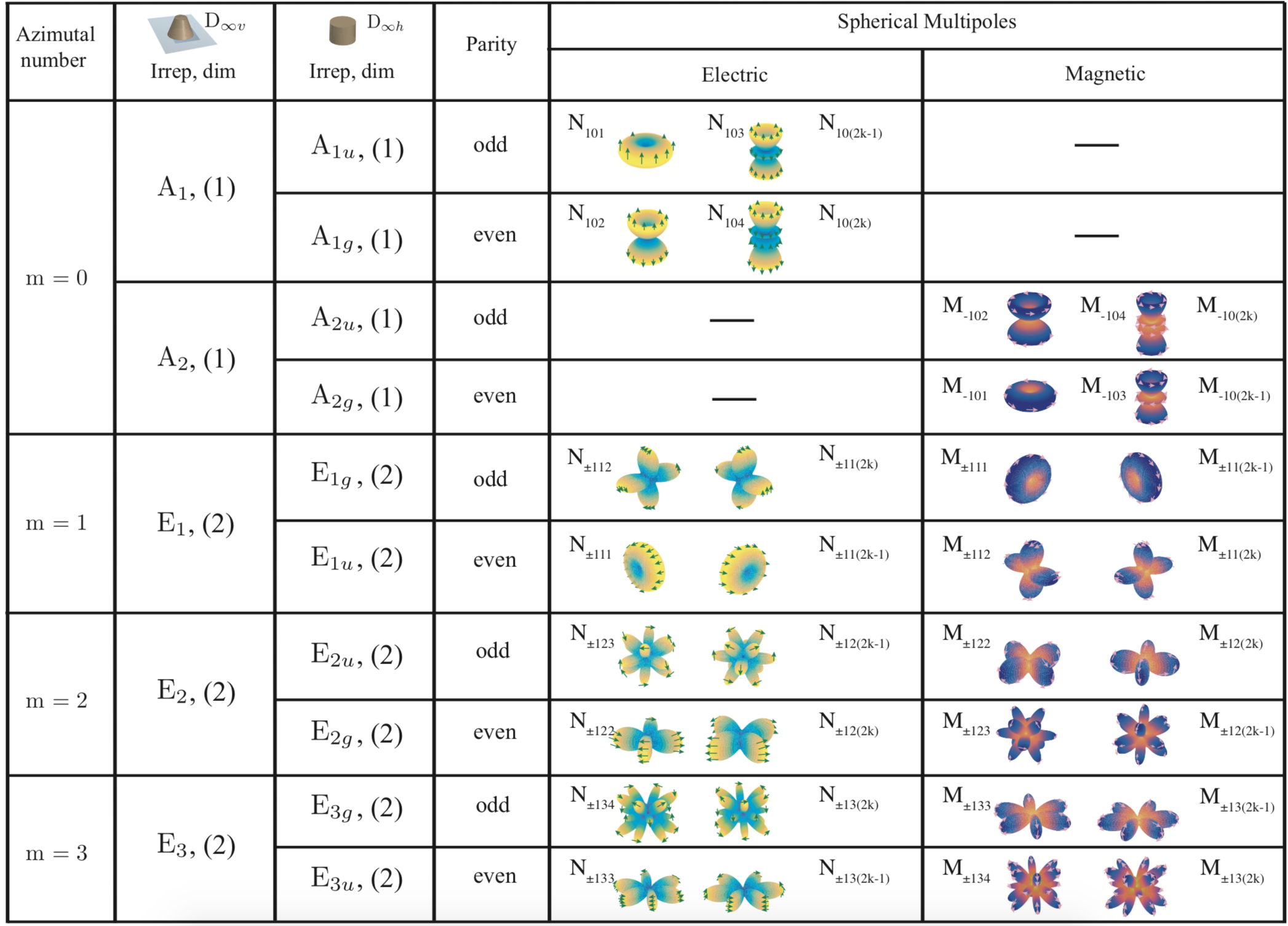}%
	\caption{ The classification of modes by  representations and the multipole composition of eigenmode for cone ($D_{\infty v}$ group), cylinder ($D_{\infty h}$ group). For each mode type showed the dimension of irreducible representations and found specific series of VSHs $\mathbf{N}_{pml}$ and $\mathbf{M}_{pml}$. Index $l$ expressed in $k$, which is a positive integer. }
	\label{cylinder}
\end{figure*}

  As an illustrative example, we consider a resonator with $D_{\infty h }$ symmetry. It can be a micro-pillar, dimer, or a simple cylindrical resonator in a homogeneous environment [see Fig.~\ref{perp}(a)]. For instance, we will consider a cylindrical resonator. In virtue of the axial symmetry, this group has infinite number of irreducible representtions corresponding to different magnetic quantum numbers $m$. Table~\ref{cylinder} shows the result, namely, the classification of the modes in resonators with $D_{\infty h }$ symmetry and the multipole content of each type of the modes. Let us show, how this table can be constructed using the proposed algorithm.      
 
\begin{enumerate} 
\item In virtue of the symmetry of $\Delta\varepsilon$ for the cylindrical resonator, the invariant functions $\psi_{p''m''l''}$ should be even under inversion and %with respect to the reflection in the $xy$-plane, symmetric with respect to the reflection $xz$-plane,  
and invariant with respect to the rotation by an arbitrary angle around the z-axis. Therefore, $p''=1$, $m''=0$, and $l''=2k$, where $k$ is a positive integer. 

\item Now let us consider, for example, harmonic $\vec{W}_{-1 1 0 1}=\mathbf{N}_{101}$ corresponding to the electric dipole oriented along the $z$-axis. Therefore, $t=-1$, $p=1$, $m=0$, and $l=1$. 

\item According to the proposed algorithm, this harmonics can be coupled only to the harmonics with $t'=-1$, $p'=1$, and $m'=0$. It follows from inequality $|1-l'|\leqslant 2k \leqslant 1+l'$ that $l'$ is an arbitrary positive integer. However, all the harmonics $\mathbf{W}_{-110l'}=\mathbf{0}$, when $l'$ is even. Therefore, all electrical harmonics $\mathbf{N}_{10(2k-1)}$ are coupled in the resonators of $D_{\infty h}$ symmetry and they do not couple to any other harmonics. 
\end{enumerate}
 All harmonics $\mathbf{N}_{10(2k-1)}$ form a basis of irreducible representation $A_{1u}$ of $D_{\infty h}$ symmetry group (see table~\ref{cylinder}) ~\cite{ohtaka1996photonic, cylinder,gelessus1995multipoles}. Therefore, if one of these harmonics contributes to a mode, then all the rest also contribute to it. Thus, there is no need to iterate over these functions again during further analysis.

Let us repeat the procedure taking function $\vec{W}_{1 -1 0 1}=\mathbf{M}_{-101}$ for the first step. Thus,  $t=1$, $p=-1$, $m=0$, and $l=1$.  This multipole corresponds to the magnetic dipole oriented along the $z$-axis of the cylinder and it is often associated with the fundamental modes of the resonators made of high-refractive-index materials~\red{\cite{kivshar2017meta}}. Following the algorithm, we can find that $t'=1$, $p'=-1$, and $m'=0$. Again, it follows from inequality $|1-l'|\leqslant 2k \leqslant 1+l'$ that $l'$ is an arbitrary positive integer. However, all the harmonics $\mathbf{W}_{1-10l'}=\mathbf{0}$, when $l'$ is even. Therefore, all magnetic harmonics $\mathbf{M}_{10(2k-1)}$ are coupled. They form a basis of irreducible representation $A_{2g}$ of $D_{\infty h}$ symmetry group (see table ~\ref{cylinder}).

Repeating the procedure for other harmonics with $m=0$, it is possible to show that magnetic harmonics $\mathbf{M}_{-10(2k)}$ are coupled and they form a basis of irreducible representation $A_{2u}$. The electric harmonics $\mathbf{N}_{10(2k)}$ are also coupled and they form a basis of  irreducible representation $A_{1g}$. Therefore, one can see that all the modes with $m=0$ in resonators of  $D_{\infty h}$ symmetry group are divided into four independent groups contributing by different multipoles. This division has the simple origin. It is well-known that for $m=0$, the solutions of Maxwell's equations can be divided into different polarizations usually denoted as TE and TM\red{~\cite{gorodetsky1994high}}. In addition, the modes of each polarization can be divided into the odd and even with respect to reflection in the $xy$-plane:
\begin{equation}
\left[
\begin{aligned}
E_\rho (\rho,\phi,-z) \\
E_\phi (\rho,\phi,-z) \\
E_z (\rho,\phi,-z)
\end{aligned}
\right]=\sigma_z
\left[
\begin{aligned}
E_\rho (\rho,\phi,z) \\
E_\phi (\rho,\phi,z) \\
-E_z (\rho,\phi,z)
\end{aligned}
\right]
\end{equation}
If  $\sigma_z=1$ then the mode is defined as even, and if $\sigma_z=-1$ then the mode is defined as odd.

\begin{figure*}[t!]
  	\includegraphics[width=0.99\linewidth]{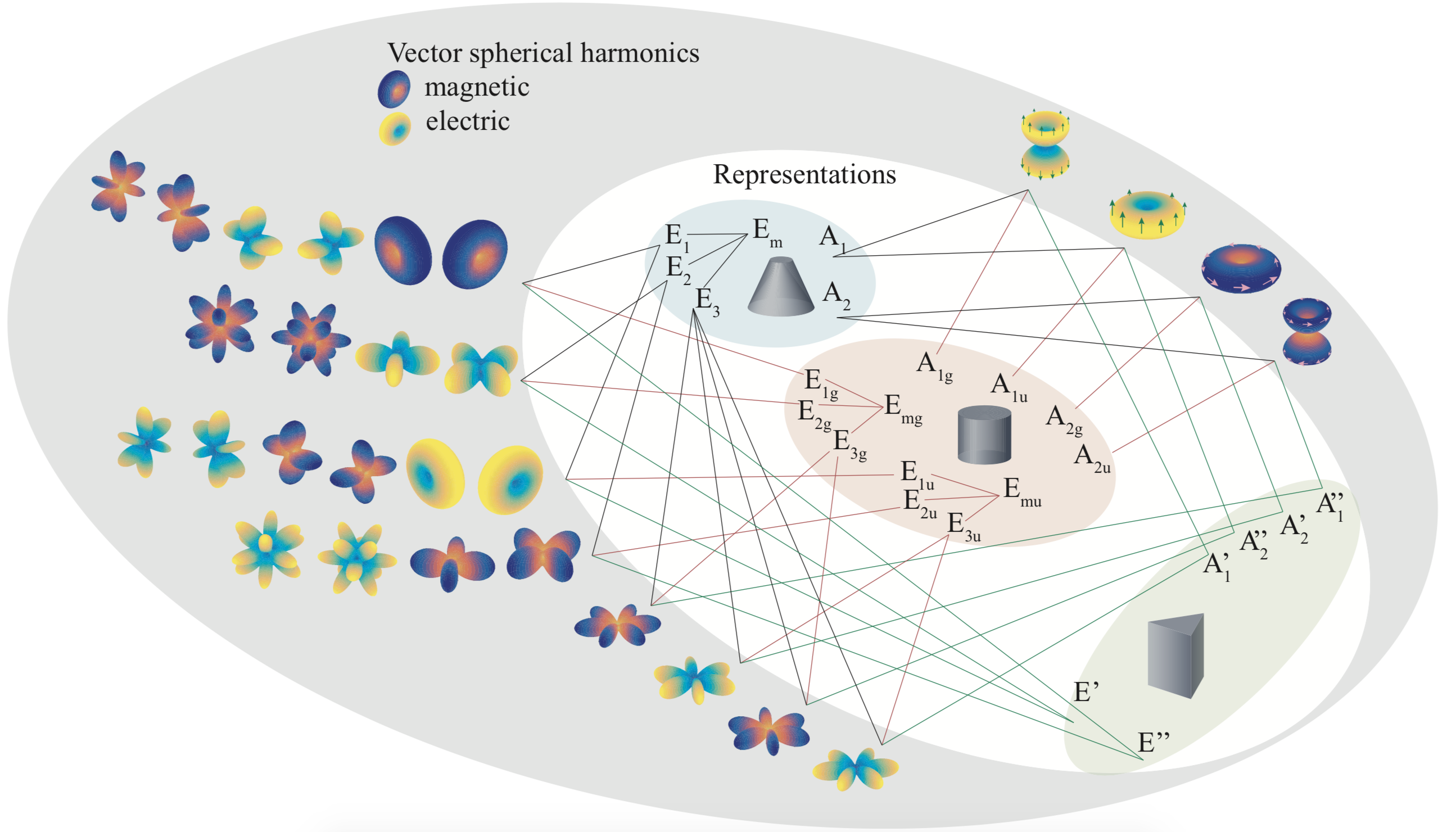}%
  	\caption{ The classification of modes by  representations and the multipole composition of each mode for cone ($D_{\infty v}$ group), cylinder ($D_{\infty h}$ group) and triangular prism ($D_{3h}$ group).}
  	\label{classific}
\end{figure*}

It follows from the proposed algorithm that only the modes with the same azimuthal index $m$ and similar parities $t$ and $p$ are coupled. However, we can note that the modes with $m\neq 0$ are transformed through each other under rotations around the $z$-axis. Indeed, for example, the $x$-oriented electric dipole ($\vec N_{111}$) turns into the $y$-oriented dipole $\vec N_{-111}$ under rotation by $\pi/2$ with respect to the $z$-axis. This means that these modes form a basis of high-dimensional irreducible representation and have the same energy. So, if under all possible transformations of the structure's symmetry the mode is only multiplied by some number (for example, $\vec N_{101}\rightarrow \vec N_{101}$  under rotations of the cylinder and $\vec N_{101}\rightarrow -\vec N_{101}$ under inversion), then the dimension of representation is one and the mode is not degenerate. If the mode transformed into some linear combination of itself and some other modes, then these modes are degenerate and the dimension of representation is more than one.  % It results in a fact that for each $m\neq0$, there are two \blue{two-dimensional} irreducible representations corresponding to odd and even modes with respect to reflection in the $xy$-plane. Because of the mixture of polarizations, the basis of both representations consists of electric and magnetic multipoles (see table~\ref{cylinder}).  

\subsection{Effect of substrate}
The classification of the modes in resonators of a cone symmetry ($D_{\infty v}$) can be straightforwardly obtained from the results for $D_{\infty h}$ symmetry group accounting for the fact that in $D_{\infty v}$ there is no mirror symmetry with respect to the $xy$-plane, and therefore, the modes are not divided into odd and even with respect to reflection in the $xy$-plane. Thus, for example, the basis of irreducible representation $A_1$ of $D_{\infty v}$ consists of all multipoles from $A_{1u}$ and $A_{1g}$ of $D_{\infty h}$, {as result, the even and odd mode are mixed} (see table~\ref{cylinder}).     

  %The cone appears from  the cylinder with reduced symmetry. Its symmetry group is $D_{\infty v}$. In this case, our definition of parity don't work and  number of types of modes decreases because of reducing the symmetry.  There is a mixing of $p_z$-even and $p_z$-odd modes in this case.

\subsection{Other symmetry groups}

  \begin{table*}[t]
 	 	\caption{ The classification of modes by  representations and the multipole composition of eigenmode for triangular prism ($D_{3 h}$). For each mode type showed the dimension of irreducible representations and found specific series of VSHs $\mathbf{N}_{pml}$ and $\mathbf{M}_{pml}$. Indices $m$ and $l$ expressed in $s$ and $k$, which are positive integer, respectively.}
\begin{tabular}{|c|c|c|c|}
	\hline
	\multirow{2}{*}{Parity} & \multirow{2}{*}{Irrep $D_{3h}$, dim} & \multicolumn{2}{c|}{Spherical Multipoles}                                          \\ \cline{3-4} 
	&                                 & Electric                                & Magnetic                                 \\ \hline
	\multirow{4}{*}{odd}    & $A_1'', (1)$                         & $\mathbf{N}_{-1,3s,2k}$                     & $\mathbf{M}_{-1,0,2k}$, $\mathbf{M}_{-1,3s,2k-1}$ \\ \cline{2-4} 
	& $A_2'', (1)$                         & $\mathbf{N}_{1,0,2k-1}$, $\mathbf{N}_{1,3s,2k}$ & $\mathbf{M}_{1,3s,2k-1}$                     \\ \cline{2-4} 
	& \multirow{2}{*}{$E'', (2)$}          & $\mathbf{N}_{\pm1,3s-1,2k-1}$               & $\mathbf{M}_{\pm1,3s-1,2k}$                  \\ \cline{3-4} 
	&                                 & $\mathbf{N}_{\pm1,3s-2,2k}$                 & $\mathbf{M}_{\pm1,3s-2,2k-1}$                \\ \hline
	\multirow{4}{*}{even}   & $A_1', (1)$                          & $\mathbf{N}_{1,0,2k}$, $\mathbf{N}_{1,3s,2k-1}$ & $\mathbf{M}_{1,3s,2k}$                       \\ \cline{2-4} 
	& $A_2', (1)$                          & $\mathbf{N}_{-1,3s,2k-1}$                   & $\mathbf{M}_{-1,0,2k-1}$, $\mathbf{M}_{-1,3s,2k}$ \\ \cline{2-4} 
	& \multirow{2}{*}{$E', (2)$}           & $\mathbf{N}_{\pm1,3s-2,2k-1}$               & $\mathbf{M}_{\pm1,3s-2,2k}$                  \\ \cline{3-4} 
	&                                 & $\mathbf{N}_{\pm1,3s-1,2k}$                 & $\mathbf{M}_{\pm1,3s-1,2k-1}$                \\ \hline
\end{tabular}
 	 	\label{Tab:3prism}
 \end{table*}
 
 \begin{table*}[t]
			\caption{The classification of modes by  representations and the multipole composition of eigenmode for quadrangular prism ($D_{4 h}$). For each mode type showed the dimension of irreducible representations and found specific series of VSHs $\mathbf{N}_{pml}$ and $\mathbf{M}_{pml}$. Indices $m$ and $l$ expressed  in $s$ and $k$, which are positive integer, respectively.}
	\begin{tabular}{|c|c|c|c|}
		\hline
		\multirow{2}{*}{Parity} & \multirow{2}{*}{Irrep $D_{4h}$, dim} & \multicolumn{2}{c|}{Spherical Multipoles}                                              \\ \cline{3-4} 
		&                                 & Electric                                  & Magnetic                                   \\ \hline
		\multirow{6}{*}{odd}    & $A_{1u}, (1)$                        & $\mathbf{N}_{-1,4s,2k-1}$                     & $\mathbf{M}_{-1,0,2k}$, $\mathbf{M}_{-1,4s,2k}$     \\ \cline{2-4} 
		& $A_{2u}, (1)$                        & $\mathbf{N}_{1,0,2k-1}$, $\mathbf{N}_{1,4s,2k-1}$ & $\mathbf{M}_{1,4s,2k}$                         \\ \cline{2-4} 
		& $B_{1u}, (1)$                        & $\mathbf{N}_{-1,4s-2,2k-1}$                   & $\mathbf{M}_{-1,4s-2,2k}$                      \\ \cline{2-4} 
		& $B_{2u}, (1)$                        & $\mathbf{N}_{1,4s-2,2k-1}$                    & $\mathbf{M}_{1,4s-2,2k}$                       \\ \cline{2-4} 
		& \multirow{2}{*}{$E_{g}, (2)$}        & $\mathbf{N}_{-1,4s-3,2k}, \mathbf{N}_{-1,4s-1,2k}$                     & $\mathbf{M}_{-1,4s-3,2k-1}, \mathbf{M}_{-1,4s-1,2k-1}$                    \\ \cline{3-4} 
		&                                 & $\mathbf{N}_{1,4s-3,2k}, \mathbf{N}_{1,4s-1,2k}$                      & $\mathbf{M}_{1,4s-3,2k-1}, \mathbf{M}_{1,4s-1,2k-1}$                     \\ \hline
		\multirow{6}{*}{even}   & $A_{1g}, (1)$                        & $\mathbf{N}_{1,0,2k}$, $\mathbf{N}_{1,4s,2k}$     & $\mathbf{M}_{1,4s,2k-1}$                       \\ \cline{2-4} 
		& $A_{2g}, (1)$                        & $\mathbf{N}_{-1,4s,2k}$                       & $\mathbf{M}_{-1,0,2k-1}$, $\mathbf{M}_{-1,4s,2k-1}$ \\ \cline{2-4} 
		& $B_{1g}, (1)$                        & $\mathbf{N}_{1,4s-2,2k}$                      & $\mathbf{M}_{1,4s-2,2k-1}$                     \\ \cline{2-4} 
		& $B_{2g}, (1)$                        & $\mathbf{N}_{-1,4s-2,2k}$                     & $\mathbf{M}_{-1,4s-2,2k-1}$                    \\ \cline{2-4} 
		& \multirow{2}{*}{$E_u, (2)$}          & $\mathbf{N}_{-1,4s-3,2k-1}, \mathbf{N}_{-1,4s-1,2k-1}$                   & $\mathbf{M}_{-1,4s-3,2k}, \mathbf{M}_{-1,4s-1,2k}$                      \\ \cline{3-4} 
		&                                 & $\mathbf{N}_{1,4s-3,2k-1}, \mathbf{N}_{1,4s-1,2k-1}$                    & $\mathbf{M}_{1,4s-3,2k}, \mathbf{M}_{1,4s-1,2k}$                       \\ \hline
	\end{tabular}
			\label{Tab:4prism}
\end{table*}

\begin{table*}[t]
		\caption{The classification of modes by  representations and the multipole composition of eigenmode for cube ($O_{ h}$). For each mode type showed the dimension of irreducible representations and found specific series of VSHs $\mathbf{N}_{pml}$ and $\mathbf{M}_{pml}$. Indices $m$ and $l$ expressed  in $s$ and $k$, which are positive integer, respectively.}
	\begin{tabular}{|c|c|c|c|}
		\hline
		\multirow{2}{*}{Parity} & \multirow{2}{*}{\begin{tabular}[c]{@{}c@{}}Irrep $O_h$, dim\end{tabular}} & \multicolumn{2}{c|}{Spherical Multipoles}                                                                           \\ \cline{3-4} 
		&                                                                                         & Electric                                                 & Magnetic                                                 \\ \hline
		\multirow{10}{*}{odd}   & $A_{1u}, (1)$                                                                                & $\mathbf{N}_{-1, 4s, 9}, \mathbf{N}_{-1, 4s, 2k+1}, k\geq 6$                                 & $\mathbf{M}_{-1, 4s, 2k}, \mathbf{M}_{-1, 0, 2k}, k\geq 2$                                   \\ \cline{2-4} 
		& $A_{2u}, (1)$                                                                                & $\mathbf{N}_{-1, 2, 3}, \mathbf{N}_{-1, 4s-2, 2k+1}, k\geq 3$                                & $\mathbf{M}_{-1, 4s-2, 2k}, k\geq 3$                                  \\ \cline{2-4} 
		& \multirow{2}{*}{$E_{u}, (2)$}                                                                & $\mathbf{N}_{-1, 4s-2, 2k+1}$                                & $\mathbf{M}_{-1, 4s-2, 2k}$                                  \\ \cline{3-4} 
		&                                                                                         & $\mathbf{N}_{-1, 4s, 2k+1}$                                  & $\mathbf{M}_{-1, 0, 2k}$, $\mathbf{M}_{-1, 4s, 2k}$              \\ \cline{2-4} 
		& $T_{1u}, (3)$                                                                                & $\mathbf{N}_{1, 4s, 2k-1}$, $\mathbf{N}_{1, 0, 2k-1}$            & $\mathbf{M}_{1, 4s, 2(k+1)}$                                 \\ \cline{2-4} 
		& $T_{2u}, (3)$                                                                                & $\mathbf{N}_{1, 4s-2, 2k+1}$                                 & $\mathbf{M}_{1, 4s-2, 2k}$                                   \\ \cline{2-4} 
		& \multirow{2}{*}{$T_{1g}, (3)$}                                                               & $\mathbf{N}_{-1, 4s-3, 2(k+1)}$, $\mathbf{N}_{-1, 4s-1, 2(k+1)}$ & $\mathbf{M}_{-1, 4s-3, 2k-1}$, $\mathbf{M}_{-1, 4s-1, 2k-1}$     \\ \cline{3-4} 
		&                                                                                         & $\mathbf{N}_{1, 4s-3, 2(k+1)}$, $\mathbf{N}_{1, 4s-1, 2(k+1)}$   & $\mathbf{M}_{1, 4s-3, 2k-1}$, $\mathbf{M}_{1, 4s-1, 2k-1}$       \\ \cline{2-4} 
		& \multirow{2}{*}{$T_{2g}, (3)$}                                                               & $\mathbf{N}_{-1, 4s-3, 2k}$, $\mathbf{N}_{-1, 4s-1, 2k}$         & $\mathbf{M}_{-1, 4s-3, 2k+1}$, $\mathbf{M}_{-1, 4s-1, 2k+1}$     \\ \cline{3-4} 
		&                                                                                         & $\mathbf{N}_{1, 4s-3, 2k}$, $\mathbf{N}_{1, 4s-1, 2k}$           & $\mathbf{M}_{1, 4s-3, 2k+1}$, $\mathbf{M}_{1, 4s-1, 2k+1}$       \\ \hline
		\multirow{10}{*}{even}  & $A_{1g}, (1)$                                                                                & $\mathbf{N}_{1, 4s, 2k}, \mathbf{N}_{1, 0, 2k}, k\geq 2$                                  & $\mathbf{M}_{1, 4s, 9}, \mathbf{M}_{1, 4s, 2k+1}, k\geq 6$                                \\ \cline{2-4} 
		& $A_{2g}, (1)$                                                                                & $\mathbf{N}_{1, 4s-2, 2k}, k\geq 3$                                   & $\mathbf{M}_{1, 2, 3}, \mathbf{M}_{1, 4s-2, 2k+1}, k\geq 3$                                 \\ \cline{2-4} 
		& \multirow{2}{*}{$E_{g}, (2)$}                                                                & $\mathbf{N}_{1, 4s-2, 2k}$                                   & $\mathbf{M}_{1, 4s-2, 2k+1}$                                 \\ \cline{3-4} 
		&                                                                                         & $\mathbf{N}_{1, 0, 2k}$, $\mathbf{N}_{1, 4s, 2k}$                 & $\mathbf{M}_{1, 4s, 2k+1}$                                   \\ \cline{2-4} 
		& $T_{1g}, (3)$                                                                                & $\mathbf{N}_{-1, 4s, 2k}$                                    & $\mathbf{M}_{-1, 0, 2k-1}$, $\mathbf{M}_{-1, 4s, 2k-1}$          \\ \cline{2-4} 
		& $T_{2g}, (3)$                                                                                & $\mathbf{N}_{-1, 4s-2, 2k}$                                  & $\mathbf{M}_{-1, 4s-2, 2k-1}$                                \\ \cline{2-4} 
		& \multirow{2}{*}{$T_{1u}, (3)$}                                                               & $\mathbf{N}_{-1, 4s-3, 2k-1}$, $\mathbf{N}_{-1, 4s-1, 2k-1}$     & $\mathbf{M}_{-1, 4s-3, 2(k+1)}$, $\mathbf{M}_{-1, 4s-1, 2(k+1)}$ \\ \cline{3-4} 
		&                                                                                         & $\mathbf{N}_{1, 4s-3, 2k-1}$, $\mathbf{N}_{1, 4s-1, 2k-1}$       & $\mathbf{M}_{1, 4s-3, 2(k+1)}$, $\mathbf{M}_{1, 4s-1, 2(k+1)}$   \\ \cline{2-4} 
		& \multirow{2}{*}{$T_{2u}, (3)$}                                                               & $\mathbf{N}_{-1, 4s-3, 2k+1}$, $\mathbf{N}_{-1, 4s-1, 2k+1}$     & $\mathbf{M}_{-1, 4s-3, 2k}$, $\mathbf{M}_{-1, 4s-1, 2k}$         \\ \cline{3-4} 
		&                                                                                         & $\mathbf{N}_{1, 4s-3, 2k+1}$, $\mathbf{N}_{1, 4s-1, 2k+1}$       & $\mathbf{M}_{1, 4s-3, 2k}$, $\mathbf{M}_{1, 4s-1, 2k}$           \\ \hline
	\end{tabular}
	  	\label{Tab:cubic}
\end{table*}

Following the proposed algorithm we classify the modes and analyzed their multipole content for the groups of the triangular prism ($D_{3h}$), quadratic prisms ($D_{4h}$), cube ($O_{h}$) and chiral resonators of $C_{4h}$ symmetry. The results of classification are shown in tables~\ref{Tab:3prism},~\ref{Tab:4prism},~\ref{Tab:cubic}, and in table~\ref{Chiral}. {Note that for $O_{h}$ and $C_{4h}$ groups we obtain that some similar vector spherical harmonics contribute to two different irreducible representations. This happens due to the fact that only one particular linear combination contributes to one mode and different combinations into others. For the case of the $O_{h}$ group, the combinations are provided in~\cite{PhysRevB.98.165110, RevModPhys.37.19}}. For the case of such resonators on a substrate, the even and odd modes should be  mixed. Alternatively to the table form the classification could be represented graphically (see Fig.~\ref{classific}).

\subsection{Bianisotropy}

\begin{figure*}[t!]
 \includegraphics[width=0.75\linewidth]{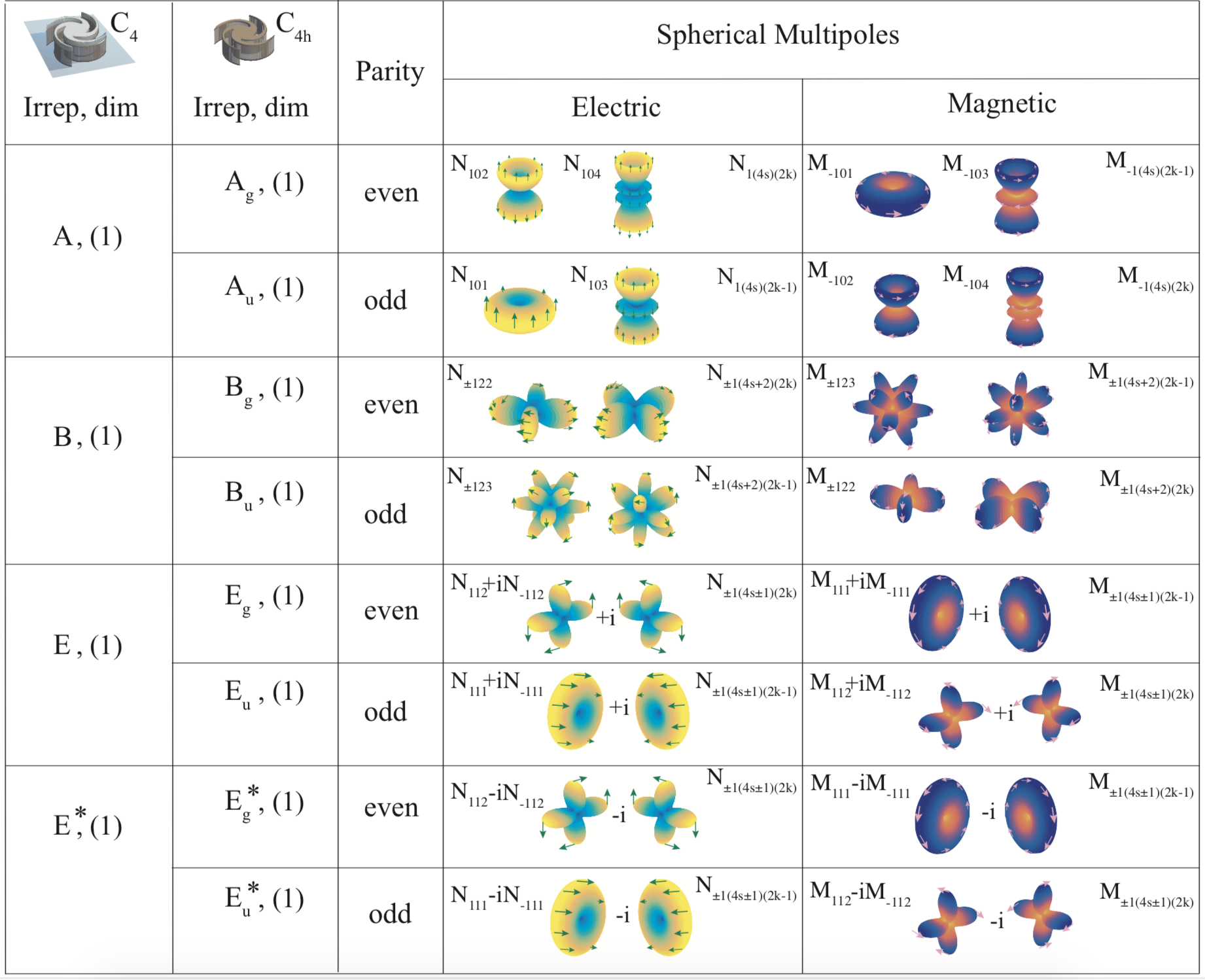}%
 \caption{The classification of modes by  representations and the multipole composition of eigenmode for chiral resonator ($C_{4 h}$ group) and chiral resonator for slab ($C_{4}$ group). For each mode type showed the dimension of irreducible representations and found specific series of VSHs $\mathbf{N}_{pml}$ and $\mathbf{M}_{pml}$. Indices $m$ and $l$ expressed in $s$ and $k$, which are positive integer, respectively. }
 \label{Chiral}
 \end{figure*}

We can observe the destruction of bianisotropy for some modes in the transition from cone ($D_{\infty v}$ group) to cylinder ($D_{\infty h}$ group), increasing the symmetry of the particle. Bianisotropy in a single particle is shown in the form of the presence of an electric and magnetic multipole of the same order simultaneously in the eigenmode ~\cite{serd}. Indeed, the presence of magnetic and electric multipoles in  one mode means, for example, that we can induce magnetic response by the incident wave with the symmetry of the electric dipole. The inversion symmetry forbids this effect. When structure has inversion symmetry, we can not obtain multipoles of different parity under inversion in one mode. Moreover, this is related not only to electric and magnetic multipoles of the same order, but also to any two multipoles of opposite parity. % For example, if you \red{??? }transfer $E_1$-type mode is divided into $E_{1g}$-type and $E_{1u}$-type modes with different inversion behavior (see Fig.~\ref{classific}).% A similar{\red{???}} effect can be observed of chiral structures when the parity{\red{???}} is broken.

%\red{Due to the fact that modes  $E$ and $E^*$ are transformed by different irreducible representations, their resonant frequency is different,  and each of them is coupled with wave of particular circular polarization, so the transmission is different for two polarizations. WRONG? SEE PODDUBNY PAPER?} 

%\subsection{Helicity scattering for systems with discrete rotational symmetry}

%According to the \cite{Fernandez-Corbaton:13}, for the systems with rotational symmetry, along the symmetry axis, forward scattering can only be helicity preserving and backward scattering can only be helicity
%flipping. These restrictions do not exist for the degree of symmetry less than 3. 

\section{Bound state in the continiuum in the triangular prism.}

\begin{figure*}[t!]
  \includegraphics[width=0.99\linewidth]{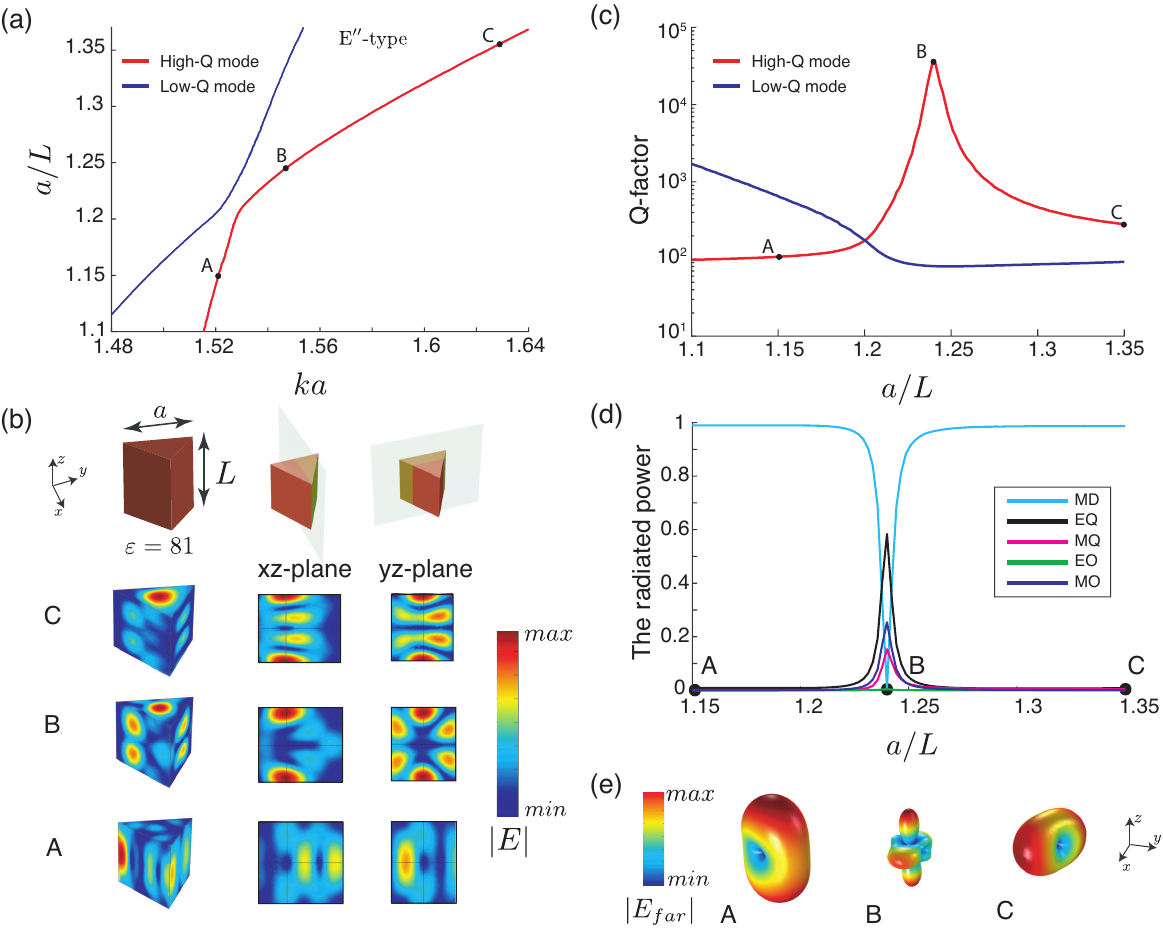}%
  \caption{(a) Strong coupling of high-Q and low-Q modes of $E''$-type in a dielectric triangular resonator with $\varepsilon=81$ depending on size parameter $ka$ and aspect ratio $a/L$, where $k,a,$ and $L$ are the wavevector in the vacuum, the side of the triangle at the base prism, and the height of the prism, respectively. (b) Distribution of the electric field amplitude $|E|$ of high-Q mode for different aspect ratios (points A, B, C). (c) Evolution of the Q-factor of two interacting modes depending on aspect ratio $a/L$. (d) Contribution of the electric and magnetic multipoles to the radiated power of high-Q mode. (e) Far-field radiation patterns for different aspect ratios (points A, B, C). Panel B corresponds to the supercavity mode.}
  \label{BIC3}
\end{figure*}

We have shown that some coefficients $C_s^q$ in expansion \eqref{eq:expansion} can be zero due to the symmetry  and, thus, some VSHs are not contribute to the modes. 
But some of the coefficients allowed by the symmetry analysis may vanish or be strongly suppressed accidentally for specific parameters of the resonators. In particular, change of the resonator's geometry preserving its symmetry can result in suppression of the radiative losses through multipole channels allowed by the symmetry. It results in a drastic increase of the quality factor (Q factor) and appearance of so-called {\it supercavity mode} or {\it quasi-bound state in the continuum} (quasi-BIC)~\cite{rybin2017high}. This effect was studied in detail theoretically and experimentally for cylindrical resonators~\cite{bogdanov2019bound,chen2018subwavelength,koshelev2020subwavelength}. Here, we show that quasi-BIC can be observed in resonators of non-cylindrical symmetry. As an example, we consider a dielectric triangular prism with a regular triangle at the base. Permittivity of the prism $\varepsilon$ is taken equal to $81$. The heigh of the prism is $L$ and the length of the base edge is $a$.

%Using the formalism described in the Chapter II , we can predict the exact multipole composition of each mode based on the symmetry group of the resonator and the irreducible representation which this mode belongs. In particular, we can better understand the nature of such a phenomenon as a quasi-bound state in a continuum (quasi-BIC).%

%A conventional device supporting light localization via a BIC-inspired mechanism is based on periodic photonic structures ~\cite{hsu2013observation} or chains of scatterers ~\cite{bulgakov2017propagating}. For these structures, strong localization can be achieved only for a large number of scatterers because it is governed by their mutual interference. Other implementations of BICs in photonic structures were presented in Refs. \cite{monticone2014embedded} and \cite{lepetit2014controlling}.%
% In the works \cite{bogdanov2019bound} and \cite{chen2018subwavelength} demonstrated that a subwavelength homogeneous dielectric resonator can support strongly interacting modes. The strong coupling regime is accompanied by the formation of a quasi-BIC when the radiative losses are almost suppressed due to the Friedrich--Wintgen destructive interference~\cite{friedrich1985interfering}.%

 Figure \ref{BIC3}(a) shows the dependence of dimensionless frequency $ka$ on aspect ration $a/L$, where $k$ is the wavenumber. The blue and red curves correspond to two eigenmodes of the resonator from the same irreducible representation $E''$. These modes have the same multipole content and they can interact through the continuum~\cite{friedrich1985interfering,cao2015dielectric}. The interaction between these modes manifests itself as a characteristic avoid crossing. As a results of the interaction, the Q factor of one mode decreases (blue curve) and for the second mode, it increases (red curve). The dependence of Q factor on the aspect ratio $L/a$ for these modes are shown in Fig.~\ref{BIC3}(c). One can see the Q factor of one of the mode are increased by more than two orders of magnitude from $1 \cdot 10^2$ to $3.6 \cdot 10^4$ [point B in Fig.~\ref{BIC3}(b)] in the vicinity of the avoid crossing, when $a/L$ changes from 1.15 to 1.25. This clearly demonstrates the appearance of quasi-BIC in non-cylindrical resonators. The distribution of the electric field amplitude for the high-Q mode for different values of the aspect ration is shown in Fig.~\ref{BIC3}(b). One can see that for the aspect ration $a/L=1.25$, when the Q factor becomes maximal, the field distribution becomes more symmetric. This indirectly indicates the suppression of scattering through the main multipole channel.

%The distributions of the  electric field of high-Q mode on the surface of the particle and in the two planes($xz$- and $yz$- planes) are shown in Fig. \ref{BIC3}(b). We observe the localization of the mode in the $yz$-plane for quasi-BIC (point B).  We have completed a multipole mode analysis for explaining the electric field localization and Q-factor growth in supercavity mode.
 
% We illustrate cancelation of radiation losses through the dominant channel in term of multipoles for $E''$-type modes.% 

 To prove that, the increase in the Q factor is the result of scattering suppression through the main multipole channel, we analyse the multipole content of the radiated field as function of the aspect ratio of the prism. It follows from the symmetry analysis (see Table~\ref{Tab:3prism}) that the lowest-order multipole contributing to the considered modes is the magnetic dipole moment ($\mathbf{M}_{\pm 1,1,1}$), next non-zero multipoles for these modes are the electric ($\mathbf{N}_{\pm 1,1,2}$) and magnetic ($\mathbf{M}_{\pm 1,2,2}$) quadropole moments, and the  electric ($\mathbf{N}_{\pm 1,2,3}$) and magnetic ($\mathbf{M}_{\pm 1,1,3}$) octupole moments. The contribution of these multipoles to the radiated power as a function of the aspect ration is shown in Fig.~\ref{BIC3}(d). The calculations were done using the COMSOL Multiphysics software package. One can see that exactly at the point B, where the Q factor reaches the maximal value, the magnetic dipole moment of the mode is suppressed and the mode behaves as electric quadrupole. It also can be seen from the far-field radiation patterns shown in Fig.~\ref{BIC3}(e).  
% 
% For the aspect ratio $a/L$ of the quasi-BIC, radiation through the magnetic dipole channel becomes negligible, the dominant radiation channel is the electric qauadropole and  magnetic octopole [see Fig. \ref{BIC3}(d)] and the radiation pattern changes dramatically in the far field [see Fig. \ref{BIC3}(e)].
% 
%
%Suppression of radiation losses through a dominant channel for quasi-BIC is possible to show using the multipole decomposition eigenmode for the  particle with any symmetry. 
%
%The Numerical calculations were carried out in a commercial package using the COMSOL Multiphysics.

\section{Conclusion}

In summary, we have classified eigenmodes of the resonators of the simplest shape using the group theory analysis. For each type of the modes we found its multipole content. The proposed algorithm can be used for the resonators of arbitrary shapes made of  any materials and placed into any environments. By the example of cylindrical and cone resonators, we demonstrated how their eigenmodes modes can be classified and how to find the specific multipole series inherit to each type of the modes. We presented ready-made tables of mode classification and their multipole content for resonators of cylindrical symmetry (on a substrate and in homogeneous environment), triangular and quadratic prisms, cube and chiral resonators. 

Many beautiful optical phenomena like anapole, invisibility, or directive light catering are usually considered only for spherical resonators because of the simplicity of their shape. However, in practice, it is difficult to have pure spherical resonators in homogeneous environment. Using the developed approach, the optical phenomena, which are explained in terms of multipole moments, can be extended beyond the resonators of spherical shape.  Thus, we predicted and explained the appearance of quasi-bound state in the continuum in the triangular prism. Therefore, the developed formalism shows the beauty and power of the symmetry analysis in physics. It allows one to engineer, predict, and explain scattering phenomena and optical properties of resonators and meta-atmos basing only on their symmetry without numerical simulations, and it can be used for the design of new photonic and microwave devices.            

\acknowledgements          

This work is supported by the Grant of the President of the Russian Federation (MK-2224.2020.2) and RFBR (19-02-00419). A.B. and K.F. acknowledge the BASIS foundation.   

%The developed approach can allows to extend  
%
%The developed approach can be applied for engineering optical properties of non-spherical resonator and     
%
%allows one to understand multipole the nature of many beuteful optical phenomena like invisibility and  directional light scattering. We    In particular we considered the supercavity mode (quasi-BIC) in a triangular prism and made a multipole decomposition thereby showing that the low dominant radiation channels are suppressed and the main contribution is made by the higher orders of multipoles.
%
%We want to highlight that theoretical group approach is very powerful method and our analysis is applicable for structures made of any materials. However, we need to mention that the provide analysis does not give absolute values of the coefficients in the multipole series. Thus, some terms can be substantially damped as happens for magnetic moments in plasmonic particles.
%

\appendix 
\section{Definition of vector spherical harmonics}
\label{dvsh}

The vector spherical harmonics are defined in the following way:
\begin{gather}
  \vec{M}_{^{-1}_{\ 1}mn}=\nabla \times (\vec r \psi_{^{\ 1}_{-1}mn})\:, \\
  \vec{N}_{^{\ 1}_{-1}mn}=\frac{\nabla \times \vec{M}_{^{-1}_{\ 1}mn}}{k}\:, \\ 
 \vec{L}_{^{ \ 1}_{-1}mn}=\nabla \psi_{^{\ 1}_{-1}mn},
\end{gather}
where
\begin{align}
 & \psi_{1mn}(kr)=z_n(kr) P_n^m(\cos\theta)\cos m\phi \:, \\
&   \psi_{-1mn}(kr)=z_n(kr) P_n^m(\cos\theta)\sin m\phi.
\end{align}
Here $z_n(kr)$ can be replaced by spherical bessel function of any kind, and $ P_n^m(\cos\theta)$ is the associated Legendre polynomial.
  
%%To make the  derivations more compact,  spherical harmonics ($\vec{N}_{p m n }$ and $\vec{M}_{p m n }$) will be denoted by the same letter $\vec{W}_{t p m n }$ with $\vec{M}_{p m n }$ denoted by index $t = (-1)^{n+1}$, and $\vec{N}_{p m n }$ by $t= (-1)^n$.  Also for $\psi_{p mn}(kr)$  $\Rightarrow$ $Y_{t p m n }$ with $\psi_{p mn}(kr)$ denoted by index  $t= (-1)^n$.
%%Index $n$ corresponds to the order of the multipole, $m$ takes values from $0$ to $n$, and $p = 1$ if the function $\vec{W}_{t p m n }$ is even with respect to reflection in the $y = 0$ -plane ($\varphi \rightarrow -\varphi$ in spherical coordinates), and $p = -1$ if it is odd. \red{Here we notice, that $\vec N$ and $\vec L$ functions with similar numbers have the similar symmetry in the respect to the O(3) group transformations~\cite{lobanov2019resonantstate}. Moreover, we will be only interested in the symmetry behavior of all these vector functions,  so for our problem we can also denote $\vec{L}_{p m n }$-harmonics  by  $\vec{W}_{t p m n }$ with $t= (-1)^n$.   }

\section{Transformation of vector functions}
\label{trans}

\begin{figure}
     	\includegraphics[width=0.95\linewidth]{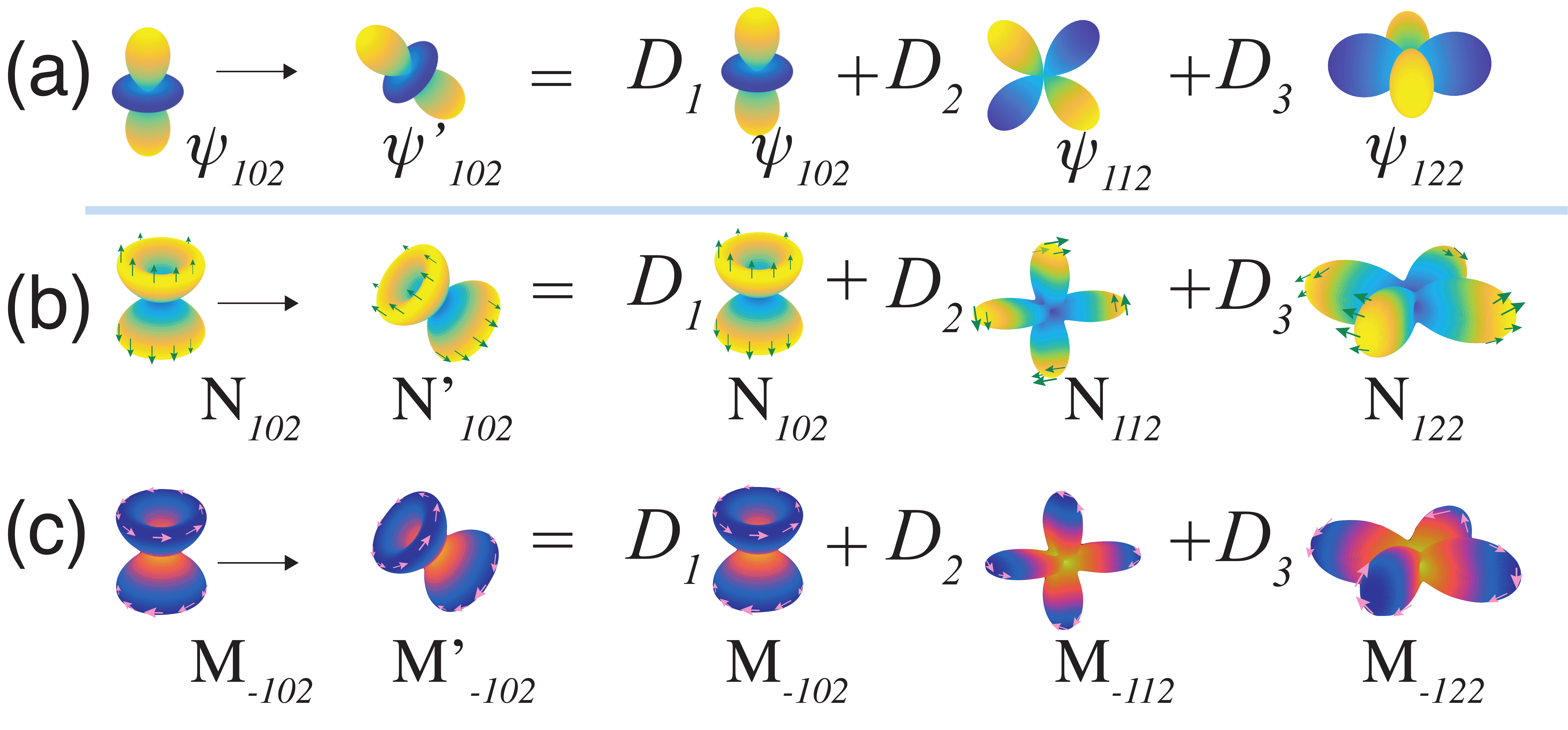} 
     	\caption{(a) Example of rotation transformation of the scalar spherical function $\psi_{102}$. The arbitrarily rotated function can be presented as linear combination of functions $Y_{^1_{-1}m2}$, and the function, rotated around the $y$-axis - as linear combination of  functions $\psi_{1m2}$. (b), (c) Rotation of electric $\mathbf{N}_{102}$ and magnetic  $\mathbf{M}_{-102}$ vector spherical harmonics is described by the similar rule, and rotated vector spherical harmonics are presented as the combination of functions $\vec W_{^1_{-1}m2}$ with the same coefficients as scalar. } 
     	\label{app1}
\end{figure}

Any point symmetry group transformation can be written as a combination of rotations and inversion, so there is no need to consider, for example, reflections, separately.
Transformation of scalar spherical functions is given by the known formula~\cite{VMK}:

\begin{gather}
\psi_{pml}\left(\hat R^{-1} \vec r\right) = \sum_{m', p'} \psi_{p'm'l}(\vec r) S_{p'pm' m}^{l}(\hat R).
\end{gather}
Here $S_{p'pm' m}^{l}(\hat R)$ are the matrices which can be obtained from the law of the transformation of complex spherical functions~\cite{Aubert2013}. They are specific combinations of the Wigner D-matrices~\cite{dwigner}, and $\hat R$ is the rotation matrix which transforms the radius vector.  Note that functions with similar $l$ are transformed through each other and the radial part $z_n(kr)$ is not transformed under point group transformations.
%For the real (tesseral) spherical functions, which can be presented as linear combinations of complex ($Y_{^{ \ m}_{-m}l}=Y_{1ml}\pm i Y_{-1ml}\propto \psi_{1ml}\pm i \psi_{-1ml}$), the rules are derived by representing them through the complex ones. 

%%\begin{gather}
%%Y_{ml}\left(\hat R^{-1} \vec r\right) = \sum_{m'} Y_{m'l}(\vec r) D_{m' m}^{l}(\hat R).
%%\end{gather}
%%Here $D_{m' m}^{l}(\hat R)$ are the {Wigner D-matrices}~\cite{dwigner}, and $\hat R$ is the rotation matrix which transforms the radius vector.  
%%For the real (tesseral) spherical functions, which can be presented as linear combinations of complex ($Y_{^{ \ m}_{-m}l}=Y_{1ml}\pm i Y_{-1ml}\propto \psi_{1ml}\pm i \psi_{-1ml}$), the rules are derived by representing them through the complex ones~\cite{Aubert2013}. 

%\begin{gather}
%\vec M_{e02}' = \vec M_{e02}  \vec N_{e22}  \vec M_{e12} (\vec r) D_{22}^{2}(\beta=\pi/2)+Y_{e02}(\vec r) D_{20}^{2}(\beta=\pi/2)
%\end{gather}

%Here D are  {Wigner D-matrixes}~\cite{dwigner}, and R is rotation matrix which transforms the radius vector.  
In case of inversion, the scalar spherical function are transformed as follows: 
%\begin{gather}
%Y_{ml}\left(- \vec r\right) =(-1)^l Y_{m'l}(\vec r) 
%\end{gather}
\begin{gather}
\psi_{pml}\left(- \vec r\right) =(-1)^l \psi_{pml}(\vec r) 
\end{gather}

One can proof that under rotations, vector spherical harmonics have similar behavior as scalar ones~\cite{Zhang:08,Stein}, which is illustrated in Fig.~\ref{app1}.   However, for the vector functions, the rotation can not be written in this simple way, because in order to rotate the vector function we need to rotate both the vector argument and the value of the function. It is convenient to write such a transformation for every projection:  
\begin{gather}
N_{x,pml}(\hat R^{-1} \vec r) \hat R^{-1} \vec{e_x}+N_{y,pml}(\hat R^{-1}\vec r) \hat R^{-1} \vec{e_y}+ \nonumber \\ + N_{z,pml}(\hat R^{-1}\vec r)  \hat R^{-1} \vec{e_z} = \sum_{m',p'} S_{p'pm'm}^l(\hat R) \vec N_{p'm'l}(\vec r)
\end{gather}
\begin{gather}
M_{x,-pml}(\hat R^{-1} \vec r) \hat R^{-1} \vec{e_x}+M_{y,-pml}(\hat R^{-1}\vec r) \hat R^{-1} \vec{e_y}+ \nonumber \\ + M_{z,-pml}(\hat R^{-1}\vec r)  \hat R^{-1} \vec{e_z} = \sum_{m',p'} S_{p'pm'm}^l(\hat R) \vec M_{-p'm'l}(\vec r)
\end{gather}
%%
%%\begin{gather}
%%W_{x,ml}(\hat R^{-1} \vec r) \hat R^{-1} \vec{e_x}+W_{y,ml}(\hat R^{-1}\vec r) \hat R^{-1} \vec{e_y}+ \nonumber \\ + W_{z,ml}(\hat R^{-1}\vec r)  \hat R^{-1} \vec{e_z} = \sum_{m'=-l}^l D_{m'm}^l(\hat R) \vec W_{m'l}(\vec r)
%%\end{gather}
Matrices $S_{p'pm'm}^l(\hat R)$ are the same as for scalar spherical functions. Note, that the rotation behavior of $\psi_{pml}, \mathbf{N}_{pm l},$ and $\mathbf{M}_{-pm l}$ is similar.

%%Wigner D-matrices are the same as for scalar spherical functions, $ \mathbf{N}_{m l}=\left(\mathbf{W}_{1 m l}+i \mathbf{W}_{-1 m l}\right) $ and $ \mathbf{N}_{-m l}=\left(\mathbf{W}_{1 m l}-i \mathbf{W}_{-1 m l}\right)(-1)^{m} \frac{(l-m) !}{(l+m) !} $ for electric harmonics, and for magnetic harmonics $ \mathbf{M}_{m l}=\left(\mathbf{M}_{-1 m l}+i \mathbf{M}_{1 m l}\right) $ and $ \mathbf{M}_{-m l}=\left(\mathbf{M}_{-1 m l}-i \mathbf{M}_{1 m l}\right)(-1)^{m} \frac{(l-m) !}{(l+m) !} $ where coefficient is appeared because of the definition of associatied Legendre polynomials with negative $m$. Note, that rotation behavior of $\psi_{pml}, \mathbf{N}_{pm l},$ and $\mathbf{M}_{-pm l}$ is similar.

Under inversion, the behavior of magnetic and electric harmonics is opposite. Indeed, for electric harmonics, it is similar the behavior of the scalar functions):
\begin{gather}
-N_{x,pml}(- \vec r) \vec{e_x}-N_{y,pml}(-\vec r) \vec{e_y}- N_{z,pml}(-\vec r) \vec{e_z}=  \nonumber  \\ = - \vec N_{pml}(-\vec r) = (-1)^l \vec N_{pml}(\vec r).
\end{gather}
Magnetic harmonics are transformed under inversion as follows:
\begin{gather}
-M_{x,pml}(- \vec r) \vec{e_x}-M_{y,pml}(-\vec r) \vec{e_y}- M_{z,pml}(-\vec r) \vec{e_z}=  \nonumber  \\ = - \vec M_{pml}(-\vec r) = (-1)^{l+1} \vec M_{pml}(\vec r)  
\end{gather}
For example, the electric dipoles are odd under inversion, electric quadrupoles are even, magnetic dipoles are even, magnetic quadrupoles are odd, and so on. 

\begin{figure*}[t!]
  	\includegraphics[width=0.79\linewidth]{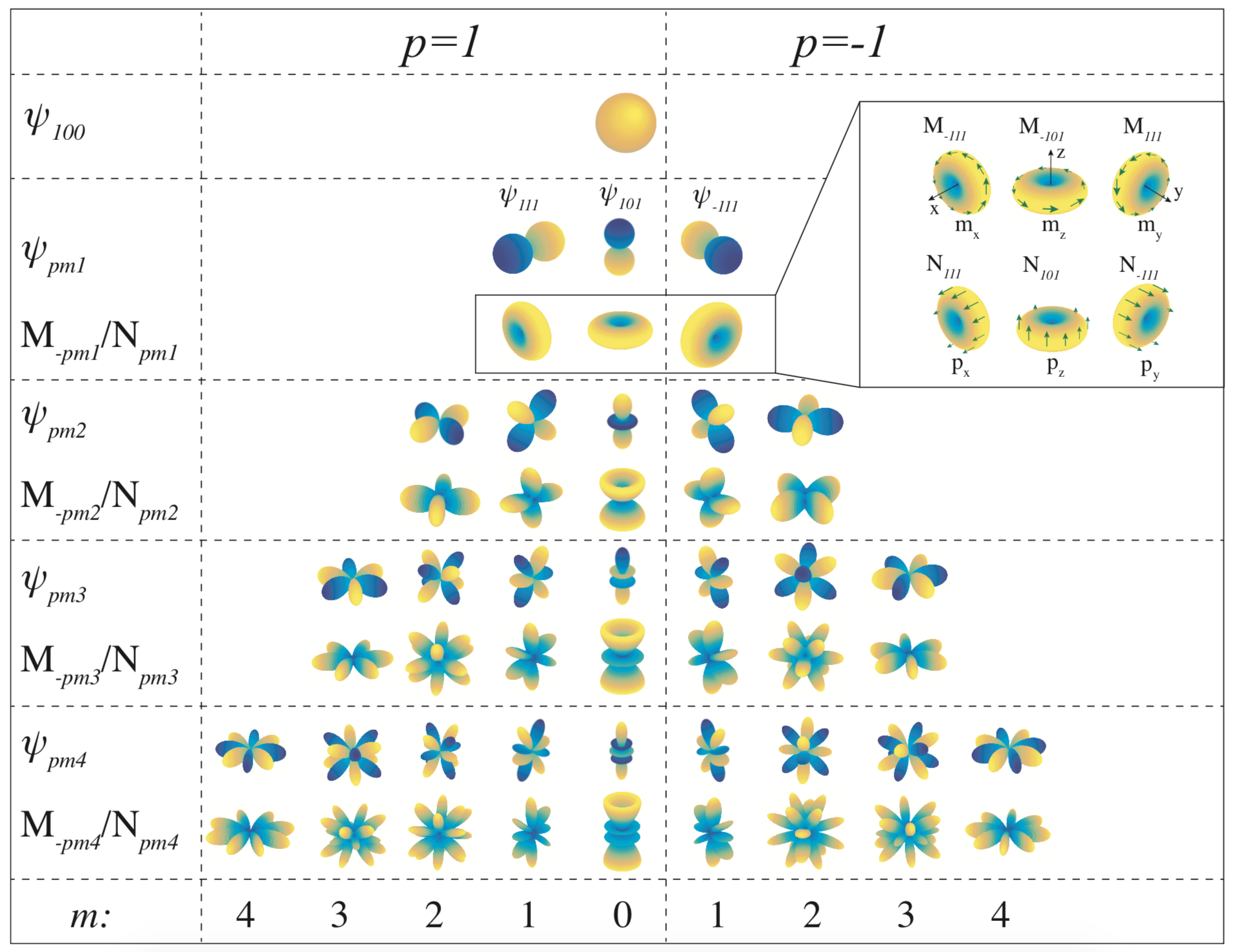}%
  	\caption{Scalar and vector real spherical harmonics for $n=1, 2, 3, 4$. Only far-field radiation patterns are presented for the vector harmonics. Radiation patterns are similar for the harmonics $\vec{N}_{pml}$ and $\vec{M}_{-pml}$, but polarization is orthogonal (example for the dipoles is in the insert).  }
  	\label{fig:harmonics}
\end{figure*}

\newpage
\bibliography{reff_abbr}

\end{document}